\documentclass[aps,prb,twocolumn,showpacs,superscriptaddress]{revtex4-1}
\usepackage{graphicx}  
\usepackage{dcolumn}   
\usepackage{bm}        

\begin{document}

\title{Superconducting properties of novel BiSe$_{2}$-based layered LaO$_{1-x}$F$_{x}$BiSe$_{2}$ single crystals}

\author{Jifeng Shao}
\affiliation{High Magnetic Field Laboratory, Chinese ACademy of Science and University of Science and Technology of China, Hefei 230026, People's Republic of China}
\author{Zhongheng Liu}
\affiliation{High Magnetic Field Laboratory, Chinese ACademy of Science and University of Science and Technology of China, Hefei 230026, People's Republic of China}
\author{Xiong Yao}
\affiliation{High Magnetic Field Laboratory, Chinese ACademy of Science and University of Science and Technology of China, Hefei 230026, People's Republic of China}
\author{Lei Zhang}
\affiliation{High Magnetic Field Laboratory, Chinese ACademy of Science and University of Science and Technology of China, Hefei 230026, People's Republic of China}
\author{Li Pi}
\affiliation{High Magnetic Field Laboratory, Chinese ACademy of Science and University of Science and Technology of China, Hefei 230026, People's Republic of China}
\author{Shun Tan}
\affiliation{High Magnetic Field Laboratory, Chinese ACademy of Science and University of Science and Technology of China, Hefei 230026, People's Republic of China}
\author{Changjin Zhang}
\email{zhangcj@hmfl.ac.cn}
\affiliation{High Magnetic Field Laboratory, Chinese ACademy of Science and University of Science and Technology of China, Hefei 230026, People's Republic of China}
\author{Yuheng Zhang}
\affiliation{High Magnetic Field Laboratory, Chinese ACademy of Science and University of Science and Technology of China, Hefei 230026, People's Republic of China}

\date{submitted to EPL on May 26, 2014, revised version received July 10, 2014, accepted by EPL on July 16, 2014}

\begin{abstract}

F-doped LaOBiSe$_{2}$ superconducting single crystals with typical size of 2$\times$4$\times$0.2 mm$^{3}$ are successfully grown
by flux method and the superconducting properties are studied. Both the superconducting transition temperature
and the shielding volume fraction are effectively improved with fluorine doping.
The LaO$_{0.48}$F$_{0.52}$BiSe$_{1.93}$ sample exhibits zero-resistivity at 3.7 K, which is higher than that of the
LaO$_{0.5}$F$_{0.5}$BiSe$_{2}$ polycrystalline sample (2.4K). Bulk superconductivity is confirmed by a clear specific-heat jump
at the associated temperature. The samples exhibit strong anisotropy and the anisotropy parameter is about 30, as
estimated by the upper critical field and effective mass model
\end{abstract}
\pacs{Valid PACS appear here}
\maketitle


\section{\label{sec:level1}Introduction}

Materials with layered structure have been intensively studied as a promising approach to the exploration of
new high transition temperature superconductors, since the discovery of cuprate superconductors\cite{1}.
This scheme has been accelerated and become much more fruitful by the discovery of the iron-based superconductors\cite{2}
and the following tremendous research. Both the cuprate and iron-based superconductors have a layered structure consisting of
the so-called superconducting layers (CuO$_{2}$ layer, Fe$_{2}$M$_{2}$(M=As, P, S, Se or Te)layer) and
the blocking charge reservoir layers\cite{3,4}. Superconductivity occurs when charge carriers are generated in the charge reservoir layers
and transferred into the superconducting layers. By changing the blocking layers or intercalating some molecular space layers,
a lot of derivative superconductors with the same superconducting layers are discovered\cite{5,6}.

Recently, novel BiS$_{2}$-based superconductivity has been reported in layered compound Bi$_{4}$O$_{4}$S$_{3}$ \cite{7},
which is composed of Bi$_{2}$O$_{2}$(SO$_{4}$)$_{(1-x)}$ (x=0.5) blocking layers and BiS$_{2}$ superconducting layers.
Subsequently, an analogous series of BiS$_{2}$-based superconductors are discovered, including
ReO$_{1-x}$F$_{x}$BiS$_{2}$ (Re= La, Ce, Nd, Yb, Pr)\cite{8,9,10,11,12}, La$_{1-x}$M$_{x}$OBiS$_{2}$ (M= Ti, Zr, Hf, Th)\cite{13}, and Sr$_{1-x}$La$_{x}$FBiS$_{2}$\cite{14,15}, ect.
The transition temperature of the BiS$_2$-based superconductors can be as high as 10.6 K\cite{8,16}.
These reports suggest that it is possible to discover more BiS$_{2}$-derivative superconducting materials.
Very recently, superconductivity with transition temperature of 2.4 K has been reported in LaO$_{0.5}$F$_{0.5}$BiSe$_{2}$ polycrystalline sample\cite{17}.
This discovery is of great importance since it represents a novel BiSe$_{2}$-based superconducting system.
However, due to the lack of single crystal, the superconducting properties remain to be investigated.
In this paper, we report the successful growth of LaO$_{1-x}$F$_{x}$BiSe$_{2}$ single crystal samples by flux method.
The superconducting parameters are determined based on the high-quality single crystals.

\section{\label{sec:level1}Experiment}

The LaO$_{1-x}$F$_{x}$BiSe$_{2}$ single crystals were grown by flux method using a mixture of CsCl and KCl as the flux (the molar ratio
of CsCl : KCl is 5 : 3)\cite{18,19,20,21}.
The starting materials of high-purity Bi$_{2}$O$_{3}$, BiF$_{3}$, Bismuth, Lanthanum and Selenium were weighed with
nominal concentrations of LaO$_{1-x}$F$_{x}$BiSe$_{2}$ and thoroughly ground in an agate mortar. Then the flux of CsCl and KCl were added,
with the mass ratio of CsCl/KCl to raw materials of 1 : 8.
The total mixture was thoroughly ground and then sealed in an evacuated quartz tube. It was slowly heated to $800\,^{\circ}\mathrm{C}$ and
kept for 48 h followed by cooling down at a rate of 2 K/h to $560\,^{\circ}\mathrm{C}$ before the furnace was shut down.
After cooling down to the room temperature, the product was removed from the quartz tube and washed by distilled water to get the
single crystal samples.

The real compositions of the obtained single crystals were determined by an energy dispersive x-ray spectroscopy (EDX) analysis,
which was performed using Oxford SWIFT3000 spectroscopy equipped with a Si detector.
The EDX measurement were done on several pieces of single crystals randomly selected from each sample.
For each piece, at least six different points were randomly selected in the EDX measurements and the average was defined as the actual composition.
The structure of the obtained crystal was characterized by powder and single crystal x-ray diffraction with Cu-K$_{\alpha}$ at room temperature.
The temperature dependence of resistivity from 2 K to 300 K was measured by a standard four-probe method in a commercial Quantum Design
physical property measurement system (PPMS-14 T) system.
Magnetic properties were performed using a superconducting quantum interference device magnetmeter.
The specific heat measured by a thermal relaxation method from 20 K down to 1.9 K was also performed on PPMS-14 T system.

\section{\label{sec:level}Results and discussion}

Figure 1 shows the scanning electron microscopy photograph of the LaO$_{0.59}$F$_{0.41}$BiSe$_{1.92}$ sample.
The obtained single crystal samples have typical dimensions of 2$\times$4$\times$0.2 mm$^{3}$. In order to determine the actual compositions
of the samples we perform an energy-dispersive x-ray spectrometry (EDX) analysis on the as-grown single crystals.
The results are given in Table 1.
The obtained values are normalized by La = 1, and the oxygen content is defined as 1-$x$ ($x$ is the content of F),
considering the inaccuracy of oxygen by EDX measurement. No Cs, K, and Cl elements are detected in the samples. For La, Bi and Se elements,
the relative ratio agrees well with the stoichiometric ratio except for a little deficiency of Se, which is similar to the deficiency of S
in NdO$_{1-x}$F$_{x}$BiS$_{2}$ single crystals \cite{22}. It can be seen that the actual F concentration increases with increasing
nominal F content. However, the actual F content saturates at about 0.5 (for example, the actual F content is $x$=0.52 in the sample
with nominal content of $x$=0.9).
Another noticeable fact is that for the samples with nominal contents $x$=0.2 and $x$=0.3, the actual F contents are about 0.37 and 0.38,
which are significantly higher than the nominal contents. These results indicate that in the LaO$_{1-x}$F$_{x}$BiSe$_{2}$ single crystals, the preferred
F content is in the $x$=0.3-0.6 region. Similar results have also been discovered in LaO$_{1-x}$F$_{x}$BiS$_{2}$ single crystals\cite{23}.
\begin{table}
\caption{\label{tab:table1} The comparison between nominal and real compositions of different F-doped LaO$_{1-x}$F$_{x}$BiSe$_{2}$ single crystals, determined by EDX. }
\begin{ruledtabular}
\begin{tabular}{lcr}
\hline Nominal composition&Measured chemical composition\\
\hline LaO$_{0.8}$F$_{0.2}$BiSe$_{2}$ & LaO$_{0.63}$F$_{0.37}$Bi$_{1.00}$Se$_{1.92}$\\
\hline LaO$_{0.7}$F$_{0.3}$BiSe$_{2}$ & LaO$_{0.62}$F$_{0.38}$Bi$_{1.00}$Se$_{1.92}$\\
\hline LaO$_{0.5}$F$_{0.5}$BiSe$_{2}$ & LaO$_{0.59}$F$_{0.41}$Bi$_{0.99}$Se$_{1.92}$\\
\hline LaO$_{0.3}$F$_{0.7}$BiSe$_{2}$ & LaO$_{0.54}$F$_{0.46}$Bi$_{0.99}$Se$_{1.93}$\\
\hline LaO$_{0.1}$F$_{0.9}$BiSe$_{2}$ & LaO$_{0.48}$F$_{0.52}$Bi$_{1.00}$Se$_{1.93}$\\
\end{tabular}
\end{ruledtabular}
\end{table}

\begin{figure}
\includegraphics[scale=0.3]{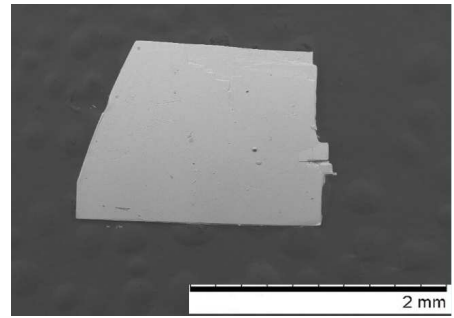}
\caption{The SEM photograph of an as-grown F-doped LaOBiSe$_{2}$ single crystal.}
\end{figure}

Figure 2(a) gives the powder x-ray diffraction (XRD) pattern of the LaO$_{0.59}$F$_{0.41}$Bi$_{0.99}$Se$_{1.92}$
sample and the refinement on the XRD pattern. The sample has a $P$4/$nmm$ tetragonal lattice with lattice constants
$a$=4.1377 \AA\ and $c$= 14.1566 \AA. The atom coordinates and site occupancy determined from the Rietveld refinement ($R_{wp}$ = 3.27\%) are also given
in Fig. 2(a). A schematic image of the crystal structure is plotted in Fig. 2(b). Figure 2(c) shows the single crystal XRD patterns of the samples.
Only (00$\ell$) diffraction peaks are observed, confirming that the samples are all single crystals and the crystallographic $c$-axis is perpendicular to the shining surface. Figure 2(d) shows the shift of the (004) diffraction peaks with increasing F content. The (004) peak
shifts toward higher angle as the F content increases from 0.38 to 0.52. Accordingly, the calculated $c$-axis lattice parameter decreases
from 14.1642 \AA\ to 14.0024 \AA. The slight shrinkage of the $c$-axis lattice parameter is reasonable
because the ionic radius of F$^{-}$ is smaller than that of O$^{2-}$.

\begin{table}
\caption{\label{tab:table2} The comparison of the lattice parameters and the transition temperatures between
the polycrystalline (P) and single crystal (S) samples of LaO$_{0.5}$F$_{0.5}$BiSe$_{2}$ and LaO$_{0.5}$F$_{0.5}$BiS$_{2}$. }
\begin{ruledtabular}
\begin{tabular}{ccccc}
\hline Compound&a (\AA)&c (\AA)& $T_{c}^{zero}$ (K)& T$_{c}$$^{susceptibility}$ (K)\\
\hline LaO$_{0.5}$F$_{0.5}$BiSe$_{2}$-P & 4.15941& 14.01567 & 2.4 & 2.6\\
\hline LaO$_{0.48}$F$_{0.52}$BiSe$_{2}$-S & 4.15663& 14.0024& 3.7 & 3.7\\
\hline LaO$_{0.5}$F$_{0.5}$BiS$_{2}$-P & 4.0527& 13.3237& 2.5& 2.7 \\
\hline LaO$_{0.54}$F$_{0.46}$BiS$_{2}$-S & \ldots& 13.39 & 3.0 & 3.0\\
\end{tabular}
\end{ruledtabular}
\end{table}

\begin{figure}
\includegraphics[scale=0.3]{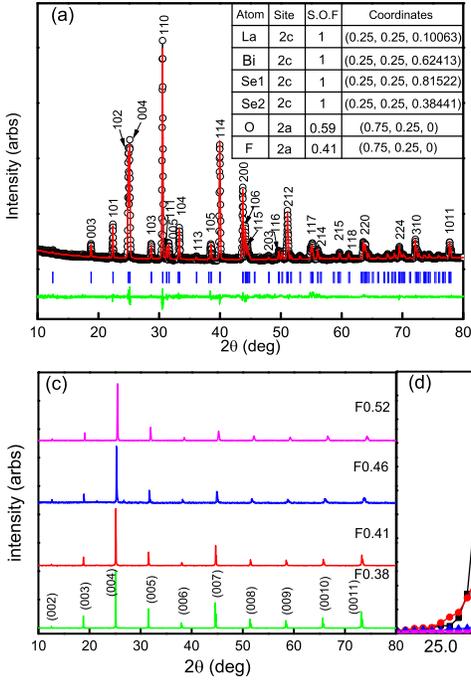}
\caption{(color online) (a) X-ray diffraction patterns of different F-doped LaO$_{1-x}$F$_{x}$BiSe$_{2}$ single crystals. (b) XRD pattern of (004) diffraction peaks for LaO$_{1-x}$F$_{x}$BiSe$_{2}$ single crystals. (c) Crystal structure of LaO$_{0.5}$F$_{0.5}$BiSe$_{2}$ derived by replacement of S by Se in the known LaO$_{0.5}$Bi$_{0.5}$S$_{2}$ structure. The orthogon indicates the unit cell.}
\end{figure}

Figure 3(a) shows the temperature dependence of resistivity of the LaO$_{1-x}$F$_{x}$BiSe$_{2}$ single crystals.
The samples exhibit a metallic-like behavior at normal state, which is consistent with the LaO$_{0.5}$F$_{0.5}$BiSe$_{2}$ polycrystalline samples\cite{17}.
The superconducting transition occurs at low temperature for all samples. For the $x$=0.52 sample, the $T_{c}^{onset}$
value is about 4.2 K and the $T_{c}^{zero}$ is 3.7 K, which are much higher than those of LaO$_{0.5}$F$_{0.5}$BiSe$_{2}$ polycrystalline sample.
Both the $T_{c}^{onset}$ and $T_{c}^{zero}$ values shift toward lower temperature as the F doping content decreases, which is shown in the inset of Fig. 3(a).
For all samples, the superconducting transition width is less than 0.5 K, suggesting the high quality of the single crystals.
In order to investigate the superconducting properties under magnetic field, we perform the resistivity versus temperature measurements at
different applied magnetic field. The results for $H$$\parallel$$ab$ and $H$$\parallel$$c$ are shown in Figs. 3(b) and (c), respectively.
It can be seen that the $T_{c}^{onset}$ value decreases and the superconducting transition width becomes larger with increasing applied magnetic field,
indicating that the superconducting state is suppressed by applying magnetic field. It can also be seen that the suppression effect
is much faster in the $H$$\parallel$$c$ case than that in the $H$$\parallel$$ab$ case.
To estimate the upper critical fields of the LaO$_{0.48}$F$_{0.52}$BiSe$_{1.93}$ sample along different direction, we plot the field dependence
curves of the $T_{c}^{onset}$ and $T_{c}^{zero}$ values for the $H$$\parallel$$ab$ and $H$$\parallel$$c$ cases.
The results are shown in the insets of Figs. 3(b) and (c), respectively. The irreversible field $\mu$$_{0}$$H_{irr}$(0) is estimated to be about 5.4 T
and 0.4 T for the $H$$\parallel$$ab$ and $H$$\parallel$$c$ cases, respectively. The upper critical field at zero temperature is estimated to be 29 T and 1 T
for the $H$$\parallel$$ab$ and $H$$\parallel$$c$ cases, as determined by the Werthamer-Helfand-Hohenberg (WHH) theory\cite{24}:
\begin{equation}
\mu_{0}H_{c2}(0)= -0.69T_{c}(d\mu_{0}H_{c2}/dT)_{T_{c}}.
\end{equation}
The anisotropy parameter is preliminary evaluated to be 33.3 using the equation:
\begin{equation}
\gamma_{s} = \frac{dH_{c2}^{\parallel ab}/dT}{dH_{c2}^{\parallel c}/dT}.
\end{equation}

Figure 3(d) shows the temperature dependence of magnetic susceptibility measured under zero-field-cooling (ZFC) and field-cooling (FC) processes
for the LaO$_{1-x}$F$_{x}$BiSe$_{2}$ samples. The applied field is parallel to the $ab$-plane.
Large diamagnetic signal is observed in all samples, confirming the occurrence of superconductivity. The diamagnetic signal is enhanced with increasing F content.
We calculate the shielding volume fraction (SVF) using the formula SVF = 4$\pi$$\chi$$\rho$/$H$ ($\rho$ is the mass density of the sample
and $H$ is the applied magnetic field). For the LaO$_{0.48}$F$_{0.52}$BiSe$_{1.93}$ sample, the estimated SVF is almost 100\% at 2K,
confirming the occurrence of bulk superconductivity. Table 2 gives a comparison on the lattice constants and the superconducting transitions
between the polycrystalline and single crystal of LaO$_{0.5}$F$_{0.5}$BiSe$_{2}$ and LaO$_{0.5}$F$_{0.5}$BiS$_{2}$ samples. It can be seen that
the single crystal samples have higher supercondusting transition temperature.
\begin{figure}
\includegraphics[scale=0.3]{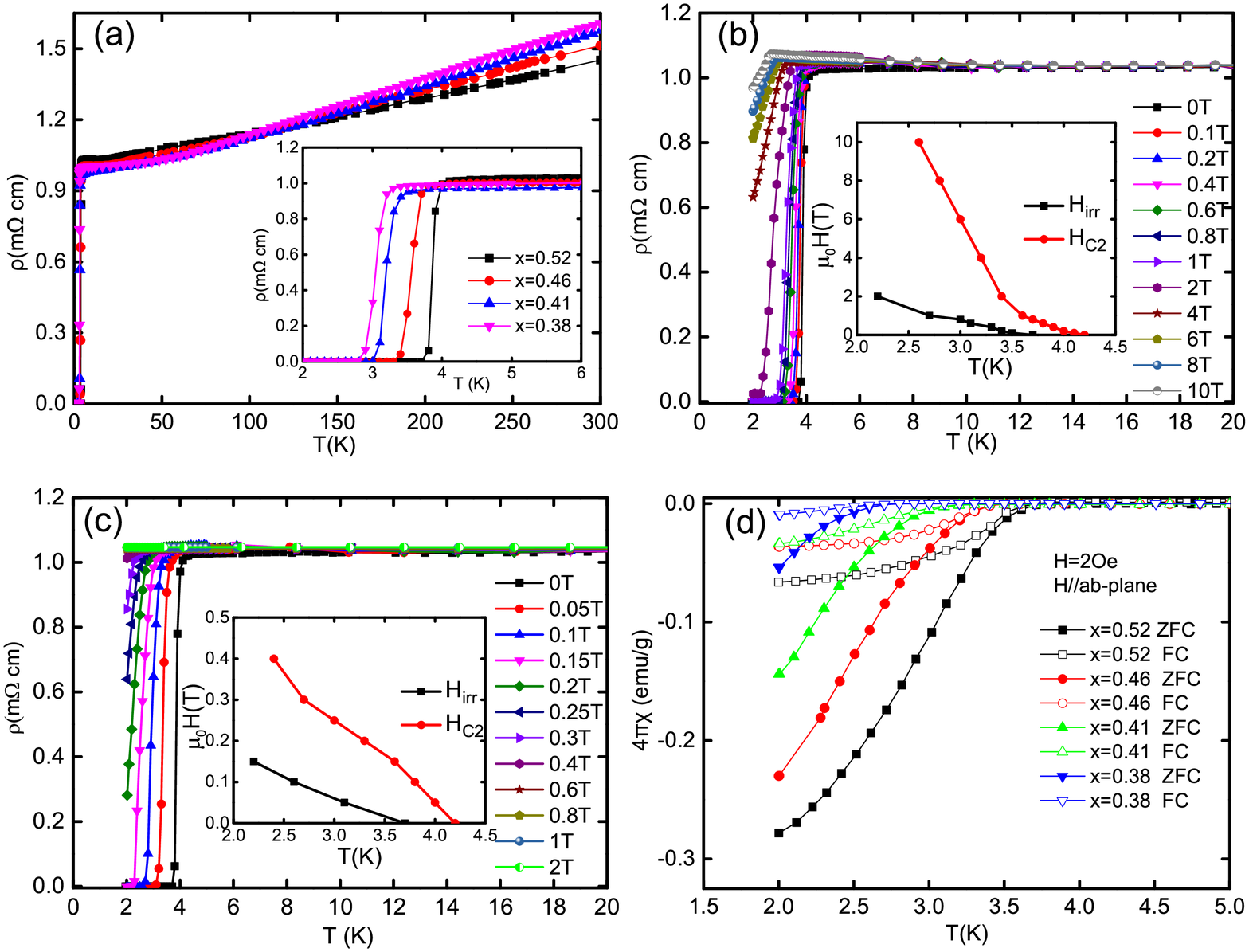}
\caption{(color online) (a) Temperature dependence of resistivity of LaO$_{1-x}$F$_{x}$BiSe$_{2}$ single crystals from 2K to 300K. The inset shows an enlarge of superconducting transition.(b)and(c) Temperature dependence of resistivity of LaO$_{0.48}$F$_{0.52\pm0.01}$Bi$_{1.00\pm0.01}$Se$_{1.93\pm0.03}$ single crystal under zero and different magnetic fields of (b)H$\parallel$ab-plane and(c) H$\parallel$c-axis. The inset of (b)and(c) shows temperature dependence of the upper critical field H$_{c2}$, and H$_{irr}$, determined from T$_{c}^{Onset}$ and T$_{c}^{Zero}$.(d) Temperature dependence of magnetic susceptibility under ZFC and FC processes for LaO$_{1-x}$F$_{x}$BiSe$_{2}$ single crystals. The applied magnetic field is 2Oe, parallel to ab-plane.}
\end{figure}

\begin{figure}
\includegraphics[scale=0.3]{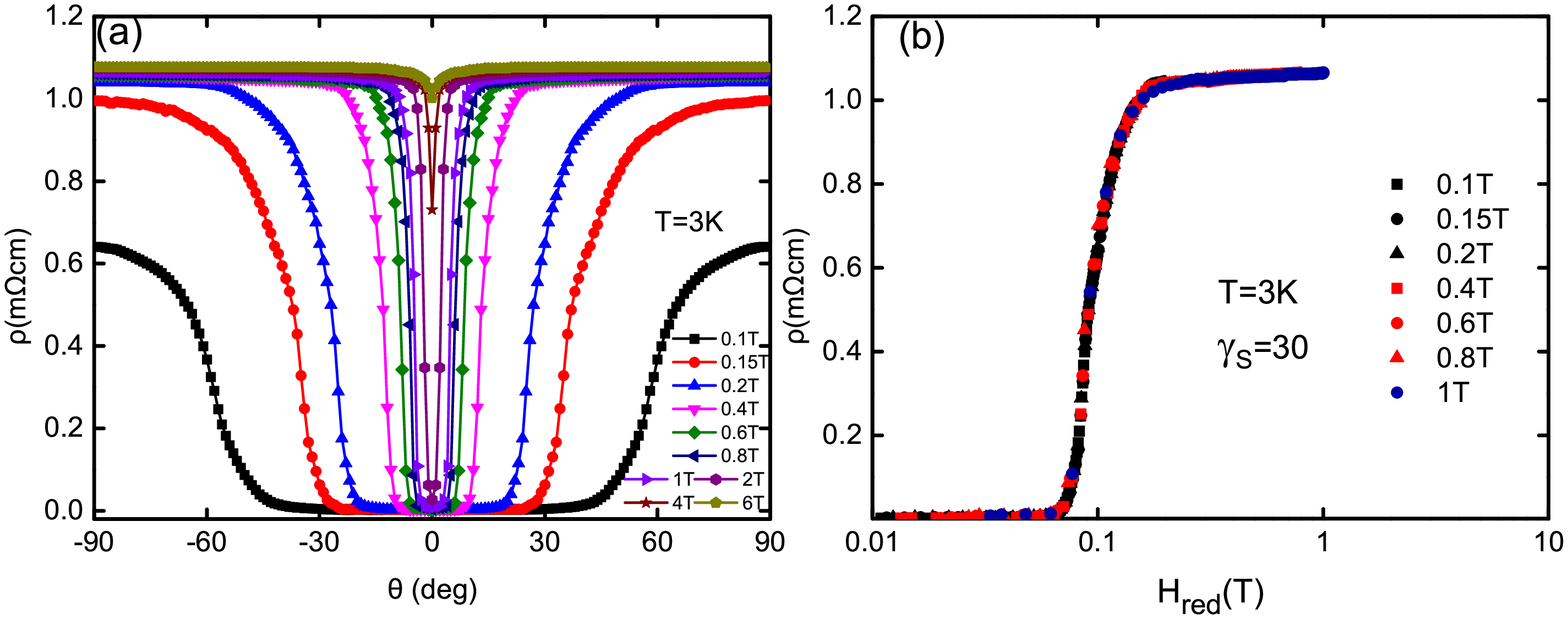}
\caption{(color online) (a) Angular dependence of resistivity of LaO$_{0.48}$F$_{0.52\pm0.01}$Bi$_{1.00\pm0.01}$Se$_{1.93\pm0.03}$ single crystal under different magnetic field from 0.1T up to 6T at fixed temperature 3K. Here, the angle $\theta$ is defined as the angle between the ab-plane and the magnetic direction. (b) Scaling of the resistivity versus the reduced magnetic field H$_{red}$ = H($\sin$$^{2}$$\theta$ + $\gamma$$_{s}$$^{-2}$$\cos$$^{2}$$\theta$)$^{1/2}$ at 3K in different magnetic field from 0.1T to 1T.}
\end{figure}

In order to scale the anisotropy parameter ($\gamma$$_{s}$) of the single crystal samples, we measure the angular ($\theta$) dependence
of resistivity of the LaO$_{0.48}$F$_{0.52}$BiSe$_{1.93}$ sample under various applied magnetic fields at a fixed temperature of 3 K. The
results are shown in Fig. 4(a). Here $\theta$ is defined as the angular between $ab$-plane and the direction of the applied magnetic field.
According to Ginzburg-Landau theory, the curves of $\rho$ versus reduced magnetic field ($H_{red}$) under various applied magnetic fields
should be fitted to one\cite{25}. The reduced magnetic field is given by $H_{red}$ = $H$($\sin$$^{2}$$\theta$ + $\gamma$$_{s}$$^{-2}$$\cos$$^{2}$$\theta$)$^{1/2}$.
As the $H_{c2}^{\parallel c}$ value for the LaO$_{0.48}$F$_{0.52}$BiSe$_{1.93}$ sample is about 1 T, we adopt the data taken from the 0-1 T range
to make the fit for $\gamma$$_s$. The result is shown in Fig. 4(b). The anisotropy parameter $\gamma$$_{s}$ at 3K is about 30,
which is close to the preliminary evaluated value of 33.3.

\begin{figure}
\includegraphics[scale=0.3]{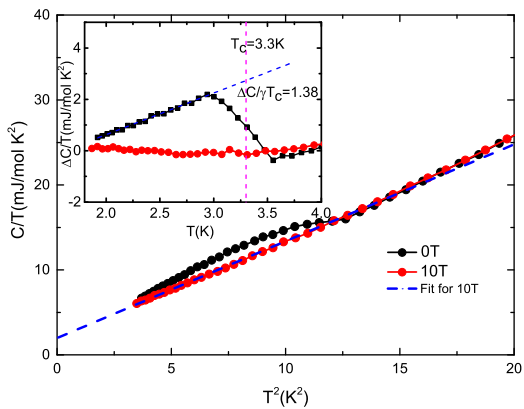}
\caption{(color online) Temperature dependence of specific heat C at superconducting state (0T) and normal state (10T) for LaO$_{0.54}$F$_{0.46\pm0.01}$Bi$_{0.99\pm0.02}$Se$_{1.93\pm0.02}$ single crystal, plotted in the form of C/T versus T$^{2}$. The dashed line is the fit to the normal state at low temperature. The inset shows temperature dependences of the difference in the electronic specific heat between the superconducting state and the normal state by subtracting the phononic contribution,C$_{e}$ = C(0T)-c(10T).}
\end{figure}

In order to confirm the occurrence of bulk superconductivity, we measure the specific heat for the LaO$_{0.54}$F$_{0.46}$BiSe$_{1.93}$
single crystal from 1.9 K to 20 K. In general, the specific heat value at low temperature is very small.
Thus we select the LaO$_{0.54}$F$_{0.46}$BiSe$_{1.93}$ sample ($T_{c}^{zero}$= 3.3 K) to perform the measurement because we can get
relatively large size of LaO$_{0.54}$F$_{0.46}$BiSe$_{1.93}$ single crystals comparing to other samples.
The temperature dependence of  specific heat in the superconducting state ($H$=0 T) and normal state ($H$=10 T) is plotted in Fig. 5.
An clear specific-heat jump occurs at about 3.3 K, confirming the occurrence of bulk superconductivity.
According to the relation of $C$/$T$=$\gamma$ + $\beta$$T^{2}$, we fit the normal state data to a polynomial. The best fitting results suggest that
the normal state electronic-specific-heat coefficient $\gamma$ and lattice coefficient $\beta$ are 1.97 mJ/mol K$^{2}$ and 1.14 mJ/mol K$^{2}$, respectively.
As the phononic contribution ($\beta$$T^{2}$) is usually independent on applied magnetic field, we get the electronic specific heat by $C_{e}$=$C$(0 T)-$C$(10 T).
The inset of Fig. 5 shows the temperature dependence of electronic specific heat in the normal and superconducting state.
The electronic specific heat jump at transition temperature (3.3 K) is evaluated to be $\Delta$C$_{e}$/$\gamma$$T_{c}$ = 1.38.
The estimated value of $\Delta$C$_{e}$/$\gamma$$T_{c}$ is comparable to the Bardeen-Cooper-Schrieffer weak-coupling limit value 1.43.
As the specific heat at the superconducting state is a little higher than that of normal state,
ultra low temperature data is needed to study the mechanism of LaO$_{1-x}$F$_{x}$BiSe$_{2}$ superconductors.

\section{\label{sec:level}Conclusion}

In summary, we report the successful growth of LaO$_{1-x}$F$_{x}$BiSe$_{2}$ superconduting single crystals for the first time.
F doping can substantially enhance the superconducting transition temperature and the shielding volume fraction.
The highest $T_{c}^{zero}$$\sim$3.7K and almost 100\% shielding volume fraction are obtained in the LaO$_{0.48}$F$_{0.52}$BiSe$_{1.93}$ sample.
The single crystal samples exhibit strong anisotropy and the
anisotropy parameter is estimated to be about 30. The upper critical fields parallel to the $c$-axis and the $ab$-plane are evaluated to be 1 T and 29 T,
respectively. We also confirm the bulk superconductivity of the LaO$_{1-x}$F$_{x}$BiSe$_{2}$ samples by specific heat measurement.

\section{\label{sec:level}Acknowledgments}

This work was supported by the State Key Project of Fundamental Research of China (Grant Nos. 2010CB923403 and 2011CBA00111), the Nature Science Foundation of China (Grant Nos. 11174290 and U1232142).


\end{document}